\documentclass[aps,amsmath]{revtex4}

\usepackage{amssymb}
\usepackage{graphicx}
\usepackage{bm}
\usepackage{color}

\begin{document}


\title{Proton Internal Pressure in $pp$ and $\bar{p}p$ Elastic Scattering}

\author{S. D. Campos} \email{sergiodc@ufscar.br}
\affiliation{Universidade Federal de S\~ao Carlos, DFQM, Sorocaba, S\~ao Paulo CEP 18052780, Brazil}


\begin{abstract}
In this work, one proposes the use of a damped confinement potential to mimic the proton internal energy. The internal pressure is calculated taking into account the occurrence of a phase transition in the $pp$ and $\bar{p}p$ total cross section. The model predicts the inversion of the positive and negative pressure regions depending on the squared energy $s_0$, where the total cross section achieves its minimum value. Considering energies below the phase transition the expected results are in accordance with the recent proton internal pressure measurement. For energies above $s_0$, the model predicts a negative pressure region near the center of the hadron surrounded by a positive pressure region.
\end{abstract}


\maketitle

\section{Introduction}\label{sec:intro}

Recently, using the deeply virtual Compton scattering, high-energy electrons were scattered from the protons in liquid hydrogen allowing the measurement of the quark pressure distribution inside the proton revealing the pressure is 10 times greater than the pressure inside a neutron star \cite{burkert_nature_vol_557_396_2018}. Moreover, there is a negative pressure region inside the proton from the periphery up to 0.6 fm, resulting the matter density is confined to a small region near the center of the hadron. Of course, the information about the matter distribution as well as the density distribution must be taken into account in the proton-proton ($pp$) and antiproton-proton ($\bar{p}p$) elastic scattering. 

A different point of view for the total cross section was recently introduced \cite{borcsik_mod_phys_lett_a31_1650066_2016}, where the negative and positive fractal dimensions for $pp$ and $\bar{p}p$ total cross section were obtained. The negative fractal dimension implies that total cross section measures the proton emptiness. On the other hand, the positive fractal dimension can be interpreted in the usual way, i.e. as an area. 

The Tsallis entropy in the impact parameter space was obtained using a few basic assumptions \cite{campos_preparation_2018}. In accordance with the phase transition observed in the total cross section, there is also a phase transition in the Tsallis entropy. The negative entropy can be seen as the internal constituents using energy to maintain its internal configuration near the center of the hadron. The positive entropy state is achieved only after a phase transition, and the internal constituents go then towards the hadron periphery.

Theoretical interpretation \cite{borcsik_mod_phys_lett_a31_1650066_2016,campos_preparation_2018} and the recent measurement in Ref. \cite{burkert_nature_vol_557_396_2018} allow the following assumption: the quarks and gluons of the hadron are located near the center, below some energy, in the elastic scattering and, as the collision energy grows, the quarks and gluons change their location towards the hadron periphery. Of course, the only constituents of the proton in this model are quarks and gluons, and the emptiness regions are fulfilled with energy due to the strong interaction. 

One proposes here the use of a damped confinement potential to mimic the hadron internal energy. Furthermore, it is necessary to introduce a phase transition factor in the internal pressure calculation in order to preserve the coherence with the theoretical approach seen in \cite{borcsik_mod_phys_lett_a31_1650066_2016,campos_preparation_2018} and with the phase transition observed in the total cross section experimental dataset. The use of both, the phase factor and the damped confinement potential, furnish an explanation for the recent measurement of the proton internal pressure \cite{burkert_nature_vol_557_396_2018}. Moreover, the pressure obtained here predicts the appearance of a gray area (hollowness effect) near the center of the hadron as the energy tends to infinity.

The paper is organized as follow. In section \ref{sec:damped} one introduces a damped confinement potential. In section \ref{sec:ip} one calculates the pressure inside a scattered hadron taking into account the phase transition occurring in the total cross section. In section \ref{sec:hollowness} are discussed the recent TOTEM and ATLAS measurement as well as the hollowness effect. Section \ref{sec:critical} presents the discussion and critical remarks.


\section{Damped Confinement Potential}\label{sec:damped}

As well-known, the confinement potential describes the potential energy of an infinitely heavy static quark-antiquark pair separated by a distance $r$. At small distances, it has the Coulomb-like behavior with the running coupling $\alpha_s(\mu)$ measuring the interaction strength. At large distances, the coupling strength is given by the string tension $\sigma$. Then, the potential exhibits its linear confining feature, i.e. the pair remains glued, and they are only free as a couple in the color singlet form. The confinement potential $V(r)$ can be explicitly written as
\begin{eqnarray}\label{eq:model_2}
V(r)=-\frac{4}{3}\frac{\alpha_s(\mu)}{r}+\sigma r,
\end{eqnarray} 

\noindent where $\alpha_{s}(\mu)$ is responsible by the strong interaction at a specific energy scale $\mu$ \cite{PDG-ChP-C40-100001-2016}. The strong increase of $\alpha_s(\mu)$ should be stopped at the infrared scale since the wavelengths of the created particles cannot exceed the size of the hadron  \cite{brodsky_phys_lett_b666_95_2008}. Explicitly, the running coupling is written in one-loop approximation as
\begin{eqnarray}\label{eq:model_3}
 \alpha_s(\mu)=\frac{1}{\beta_{0}\ln\bigl(\mu^{2}/\Lambda_{\scriptsize{\mbox{QCD}}}^{2}\bigr)},
\end{eqnarray}

\noindent where $\beta_{0} = (33-2n_{f})/12\pi$ is the $1$-loop $\beta$-function. The number of active quark flavors at the energy scale $\mu$ is given by $n_{f}$ and are considered light $m_{q} \ll \mu$, where $m_{q}$ is the quark mass: $n_f=6$ for $\mu\geq m_t$, $n_f=5$ for $m_b\leq \mu \leq m_t$, $n_f=4$ for $m_c\leq \mu \leq m_b$ and $n_f=3$ for $\mu\leq m_c$ \cite{buras_book}. Furthermore, the running coupling constant can be defined from any physical
observable perturbatively calculated \cite{grunberg_phys_lett_b95_70_1980}.

The string tension $\sigma  \approx 0.405$ GeV \cite{PRD-90-074017-2014}, in general, depends on the temperature for cold strongly interacting matter. The $\Lambda_{\scriptsize{\mbox{QCD}}}$-parameter is a non-universal scale dependent on the renormalization scheme and corresponds to the scale where the perturbatively-defined coupling would diverge \cite{PDG-ChP-C40-100001-2016}. The numerical value of $\Lambda_{\scriptsize{\mbox{QCD}}}$ depends, in particular, on $n_{f}$ and here one uses $\Lambda_{\scriptsize{\mbox{QCD}}}$ from \cite{PDG-ChP-C40-100001-2016}, for a given $n_{f}$.

As can be observed from Eq. (\ref{eq:model_2}), one requires $\mu > \mu_{\scriptsize{\mbox{min}}} \equiv \omega \Lambda_{\scriptsize{\mbox{QCD}}}$ to preserve the perturbative definition of $\alpha_s(\mu)$. The softest case corresponds to $\omega=1$ and more conservative and exact estimation is given by \cite{grunberg_phys_lett_b95_70_1980}
\begin{eqnarray}\label{eq:model_alpha_3} 
\omega =
\exp\bigl[F_{0}(\alpha_s^{\scriptsize{\mbox{max}}})/2\beta_{0}\bigr],
\end{eqnarray}

\noindent where $\alpha_s^{\scriptsize{\mbox{max}}}=\beta_{0}/\beta_{1}$,
$F_{0}(x)=x^{-1}+\beta_{1}/\beta_{0}\ln(\beta_{0}x)$, $\beta_{1}=(153-9n_{f})/24\pi^{2}$ is the $2$-loop $\beta$-function coefficient \cite{PDG-ChP-C40-100001-2016}. There are several estimation of $\mu$ based on $Y_{h}^{\scriptsize{\mbox{exp}}}$, an experimentally measurable quantity. In hadronic collisions, for instance, $\mu=Y_{h}^{\scriptsize{\mbox{exp}}}$ at $Y_{h}^{\scriptsize{\mbox{exp}}} \equiv p_{T}^{\scriptsize{\mbox{max}}}$ \cite{PRD-86-014022-2012,khachatryan_eur_phys_j_c_75_288_2015} or $Y_{h}^{\scriptsize{\mbox{exp}}} \equiv m_{3}$ \cite{EPJC-75-186-2015}, where $p_{T}^{\scriptsize{\mbox{max}}}$ is the transverse momentum of the leading jet, and $m_{3}$ is the invariant mass of the three jets leading in $p_{T}$. Here, one uses $\mu=Q^2$, the transferred momentum in the gluon frame.


There is a sea of quarks and antiquarks inside the hadron emerging from the vacuum fluctuations. It is expected the annihilation of every such created pair, remaining only the valence quarks as those which gives to the hadron its physical features. Despite pairs annihilation process, the confinement potential due to every quark-antiquark pair inside the hadron should be taken into account to produce the total potential energy inside the hadron, in a given time interval $\Delta t$. It is natural to suppose that the effective number of pairs, set at distance $r$, as being an increasing function of $Q^2$ as well as implying the enhance of $V(r)$ as $Q^2$ grows. Of course, the distance $r$ varies from some minimum up to the maximum size allowed by the system, i.e. the hadron diameter. Moreover, note that phase transition from a confining to a non-confining regime occurs at $r_0$, the root of the damped potential which is dependent on the ratio of the running coupling to string tension given by
\begin{eqnarray}\label{eq:model_2.1}
r_0=2\sqrt{\frac{\alpha_s(Q^2)}{3\sigma}},
\end{eqnarray}

\noindent and, as expected, as $Q^2$ increases this root slowly narrows to zero, and its physical meaning is quite simple: it represents the point where both terms of Eq. (\ref{eq:model_2}) contribute to $V(r)$ equally. It also can be viewed as a topological transition point in the sense that for distances $r<r_0$, the $q\bar{q}$-pair are subject to a Coulomb-like potential, i.e. they tend to behave as a free (color) charged gas. Considering $r>r_0$, they behave as a color singlet. Therefore, there is a transition point from a non-charged ($r>r_0$) to a charged gas ($r<r_0$). Furthermore, as stated above, the maximum distance of a pair corresponds to the diameter of the hadron. Therefore, the pairs set at this distance, in fact, measures a) the maximum strength of the confinement interaction and b) the hadron effective size. 

Suppose the quarks and gluons are surrounded by a sphere of radius $r$ with negative pressure above $r_0$ up to the edge of the hadron, as obtained in \cite{burkert_nature_vol_557_396_2018}. The finite size of the hadron ensures this region is bounded by a damping mechanism, acting on the confinement potential. Writing the damped confinement potential $V_{d}(r)$ as 
\begin{eqnarray}\label{eq:model_1}
V_{d}(r)=nV(r)\exp(-kr).
\end{eqnarray}

\noindent where, for the sake of simplicity, one assumes $n$ as the finite number of effective pairs at distance $r$. The wavenumber $k=1/\lambda$ can be defined in terms of $\lambda$, the Debye length and, in general, this length is a function of temperature. The factor $k$ is used here to measure the physical effects of the damped confinement potential. Furthermore, $k$ can also be seen as the Debye screening mass, $m_D$. Usually, the Debye mass is supposed as being the same for both confining and non-confining phases of $V(r)$. However, the Debye mass can be separated into two contributions: one for the confining and another one for the non-confining phase. Considering the non-confining term of Eq. (\ref{eq:model_2}), the factor $m_D$ has a linear behavior, and for the confining mechanism, it is strongly suppressed. One adopts here, for the sake of simplicity, only one wavenumber as being able to damp the confinement potential.

In the so-called Quark-Gluon Plasma (QGP) the Debye length is about $0.15$ fm \cite{burnier_phys_rev_lett_114_082001_2015}. Here, one assumes that $\lambda>0.15$ fm, i.e. the constituents are not in the QGP regime. Accordingly, one consider here $0.2\lesssim \lambda \lesssim 0.5$ fm, obtaining $2\lesssim k\lesssim 5$ fm$^{-1}$, then, for the sake of simplicity, in all figures one adopts $k=3$ fm$^{-1}$. Moreover, one observed a weak dependence on the number of active flavors $n_f$ and, thus, in all figures $n_f=3$ is fixed.

\begin{figure}
\centering{\includegraphics[scale=0.8]{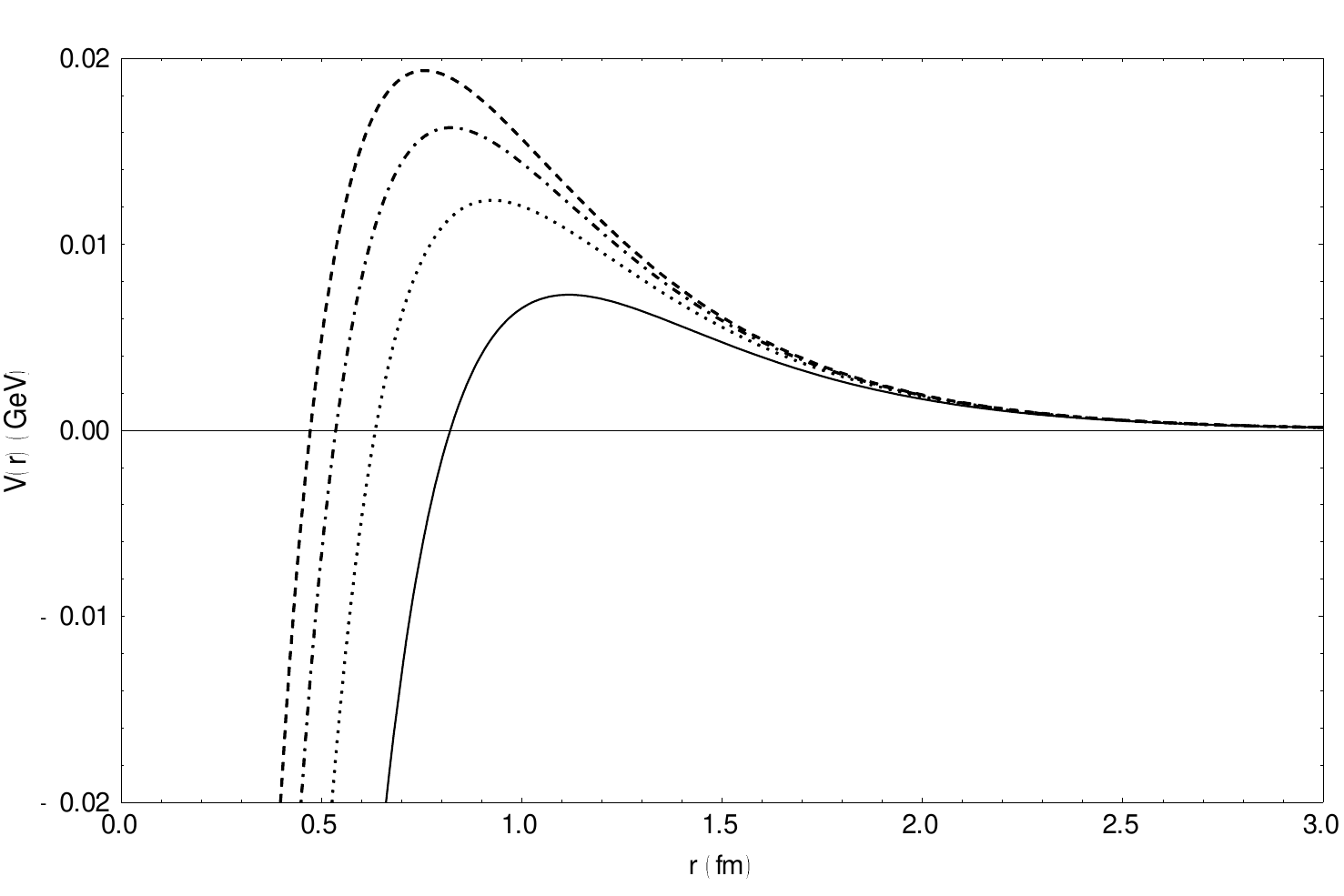}
}
\caption{Damped confinement potential for $Q^2=10^2$ GeV$^2$ (solid), $Q^2=10^4$ GeV$^2$ (dotted), $Q=10^6$ GeV$^2$ (dot-dashed) and $Q=10^8$ GeV$^2$ (dashed). For each transferred momentum one uses $k=3$ fm$^{-1}$ and $n_f=3$.}
\label{fig:conf_pot1}
\end{figure}

Figure \ref{fig:conf_pot1} shows Eq. (\ref{eq:model_1}) for different $Q^2$, $k=3$ fm$^{-1}$ and $n_f=3$. The transferred momentum $Q^2$ grows as well as the maximum value of $V_{d}(r)$, and, on the other hand, the root given by Eq. (\ref{eq:model_2.1}) shrinks. 

\section{Internal Pressure}\label{sec:ip}

The pressure inside a proton can be defined in the same way the pressure inside a gas or liquid composed of molecules. However, the proton is composed of quarks and gluons interacting via the confinement potential. The first measurement of such mechanical property is due to Burkert, Elouadrhiri, and Girod (BEG) using Generalized Parton Distributions (GPD) instead of the Gravitational Form Factors (GFF) \cite{pagels_phys_rev_144_1250_1966}, which is, in fact, the correct way to measure the pressure if the graviton is known \cite{polyakov_phys_lett_b_555_57_2003}. However, the graviton-proton scattering cannot be performed yet, and the way used by BEG to access information about the pressure was using GPD with the help of the deeply virtual Compton scattering of electrons by a proton in liquid hydrogen. In this scheme, the quark structure is probed with high-energy virtual photons exchanged between the electron and the proton, and a real photon controls the transferred momentum $t$ to the proton. The GPD is connected to the Compton Form Factor from the deeply virtual Compton scattering data and, finally, sum rules relating Mellin moments of the GFF and GPD are used to obtain the form factor $d_1(t)$, which is responsible by the description of the force and pressure distribution inside the proton. 

The pressure radial distribution, $p(r)$, inside the hadron depends on the temperature, chemical potential of quarks and gluons, for instance. One calculates here only the pressure as a result of the damped confinement potential shown by Eq. (\ref{eq:model_1}), neglecting possible contributions from kinematic terms. Of course, as the collision energy rise, the volume of the hadron also is expected to grow, achieving some effective size. The increasing energy and transferred momentum may also enhance the number of effective $q\bar{q}$-pairs. Therefore, the following constraint is stated: the ratio of the effective number of pairs created to the hadron volume is constant at every $\sqrt{s}$. Then, one considers a constant number density given by
\begin{eqnarray}\label{eq:ratio}
\frac{n}{volume~of~the~hadron}=\zeta~~(\mathrm{fm}^{-3}).
\end{eqnarray}

The total cross section of the $pp$ and $\bar{p}p$ elastic scattering shows a clear phase transition occurring at some squared energy $s_0$, where the total cross section achieves its minimum. This is a dynamical quantum phase transition since the system is in a non-equilibrium condition. Furthermore, this is a first order phase transition considering that one has a discontinuous change in the density, which is the inverse of the first derivative of the free energy with respect to pressure. When $s<s_0$, the total cross section decreases up to some minimum value at $s_0$. However, when $s>s_0$ the total cross section starts to rise as predicted by Cheng and Wu seminal papers \cite{cheng_phys_rev_lett_24_1456_1970,cheng_phys_rev_d1_1064_1970}. Moreover, as shown in Borcsik and Campos \cite{borcsik_mod_phys_lett_a31_1650066_2016}, the total cross section behavior can be explained by the use of the concept of fractal dimensions. When $s<s_0$, the fractal dimension of the total cross section is negative, measuring the emptiness of the hadron. Otherwise, if $s>s_0$ the fractal dimension is positive and the total cross section possess the usual physical interpretation. In addition, the Tsallis entropy approach in the impact parameter space \cite{campos_preparation_2018} also exhibits a peculiar behavior: if $s<s_0$, the entropy is negative implying in the hadron internal constituents self-organization. On the other hand, if $s>s_0$ the entropy is positive. Therefore, the phase transition occurring at $s_0$ should be taken into account in the calculation of any physical quantity of the elastic scattering. 

Bear in mind this phase transition, one introduces in the pressure calculation inside the proton the transition factor written as
\begin{eqnarray}\label{tfactor}
g=\frac{1}{s/s_0-1},
\end{eqnarray}

\noindent where $s_0$ is the phase transition squared energy observed in the total cross section dataset. Hence, the resulting pressure of the damped confinement potential shown by Eq.(\ref{eq:model_1}) and for a fixed-$s$ and $Q^2$, is written as 
\begin{eqnarray}\label{eq:model_5}
p(r)=g\zeta\left(-\frac{4}{3}\frac{\alpha_s(Q^2)}{r}+\sigma r\right)\exp(-kr).
\end{eqnarray}

This result has only one physical root also given by Eq. (\ref{eq:model_2.1}), and it separates the negative and positive pressure regions as shown below. The stability condition is written here as
\begin{eqnarray}\label{eq:stab}
\int_0^{\infty}r^2p(r)dr=0.
\end{eqnarray}

\begin{figure}\centering{
\includegraphics[scale=0.55]{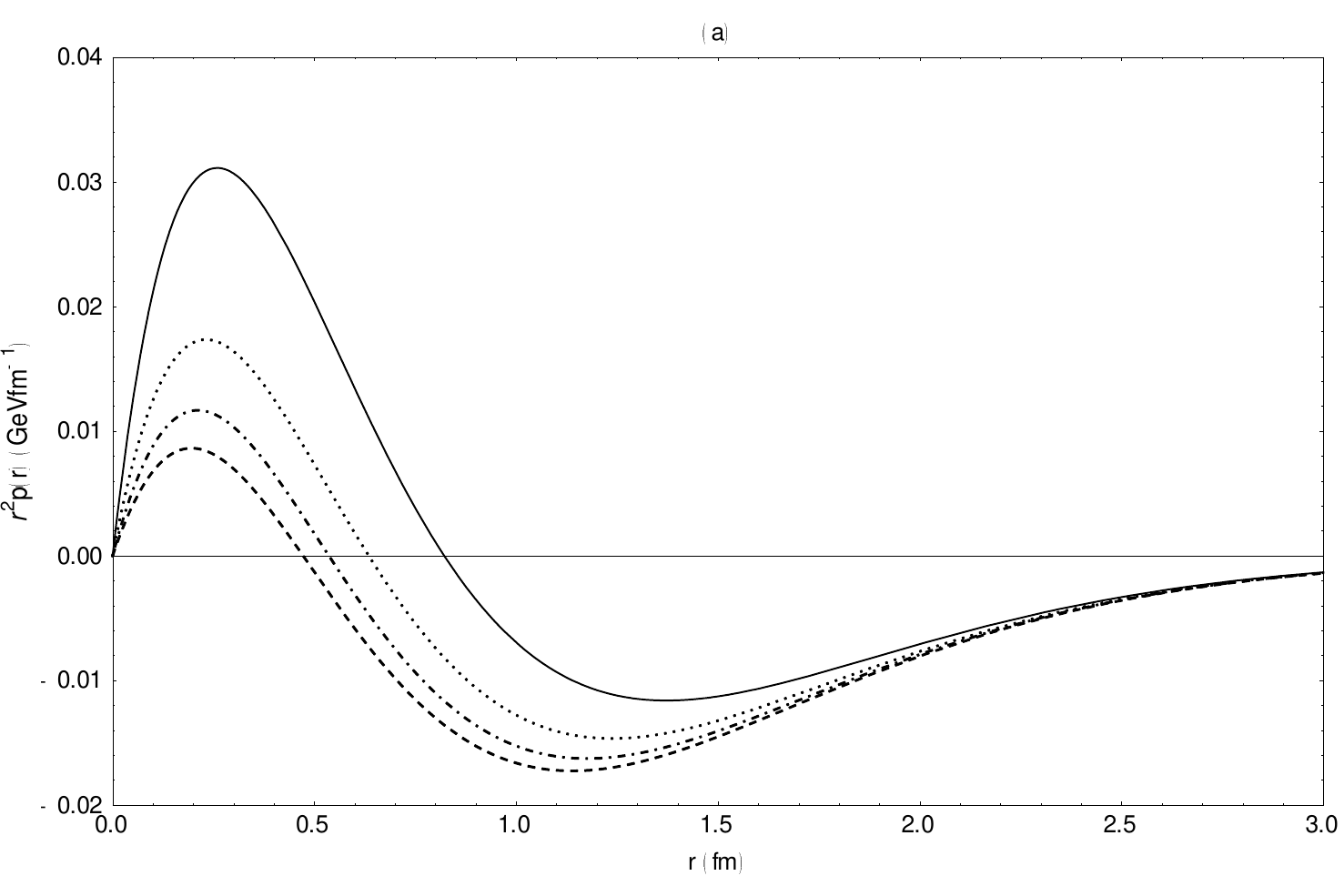}
\includegraphics[scale=0.55]{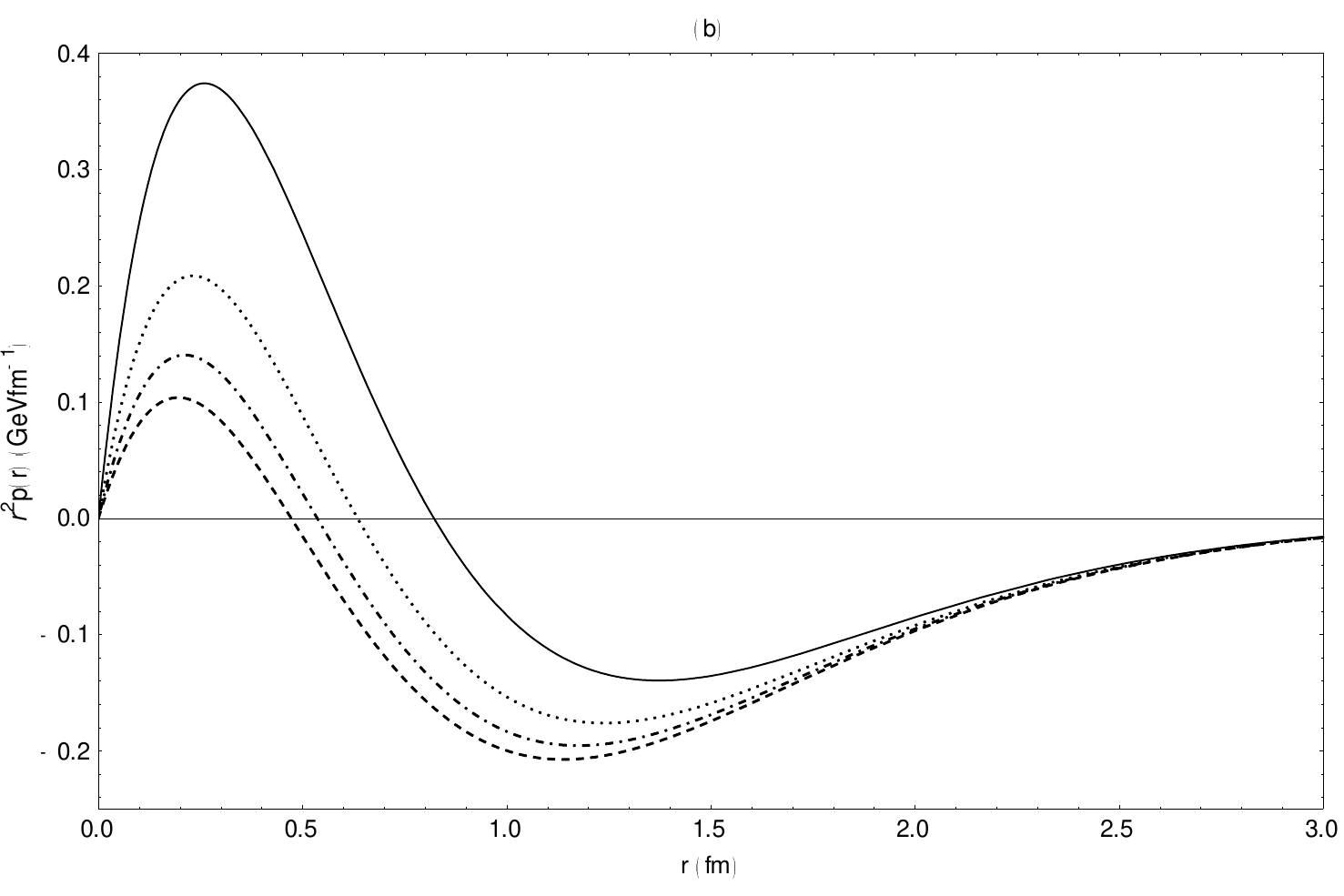}
}
\caption{Pressure distribution inside the hadron as a result from the quark interactions versus the radial distance $r$ from the center of the hadron. In all cases, $k=3$ fm$^{-1}$ and $n_f=3$. In (a) $\sqrt{s}=6$ GeV and (b) $\sqrt{s}=24$ GeV and for both cases $\sqrt{s_0}=25$ GeV and $Q^2=10^2$ GeV$^2$ (solid), $Q^2=10^4$ GeV$^2$ (dotted), $Q=10^6$ GeV$^2$ (dot-dashed) and $Q=10^8$ GeV$^2$ (dashed). The positive pressure region narrows as the $Q^2$ grows.}
\label{fig:conf_pres_s<s_0}
\end{figure}

Figure \ref{fig:conf_pres_s<s_0} shows the behavior of $r^2p(r)$ {\it versus} $r$ considering $n_f=3$ and $k=3$ fm$^{-1}$. For \ref{fig:conf_pres_s<s_0}a one uses $\sqrt{s}=6$ GeV and for \ref{fig:conf_pres_s<s_0}b $\sqrt{s}=24$ GeV. In both cases, the negative pressure region is located at the periphery of the hadron shielding a core containing pairs subject to the Coulomb-like potential since the pairs are created in the region $r<r_0$. Moreover, the core narrows and its peak shrinks as $Q^2$ grows. On the other hand, as the energy increases the peaks of $|p(r)|$ also increases. It is interesting to compare this result with those recently obtained \cite{burkert_nature_vol_557_396_2018}. The negative pressure region settles down between 0.5 and 1.0 fm and does not depends on the energy of the collision but depends strongly on $Q^2$, since the root depends on $Q^2$. Moreover, the negative pressure region grows as the collision energy tends to $\sqrt{s_0}=25$ GeV. Therefore, one can predict here two main results for future experiments as carrying out by Burkert \textit{et al.} \cite{burkert_nature_vol_557_396_2018}. The first one is the growth of $|p(r)|$ in the negative region as the energy tends to $\sqrt{s}=25$ GeV. The second one is the shrinkage of the positive pressure region toward the center of the hadron as $Q^2$ grows.

\begin{figure}\centering{
\includegraphics[scale=0.55]{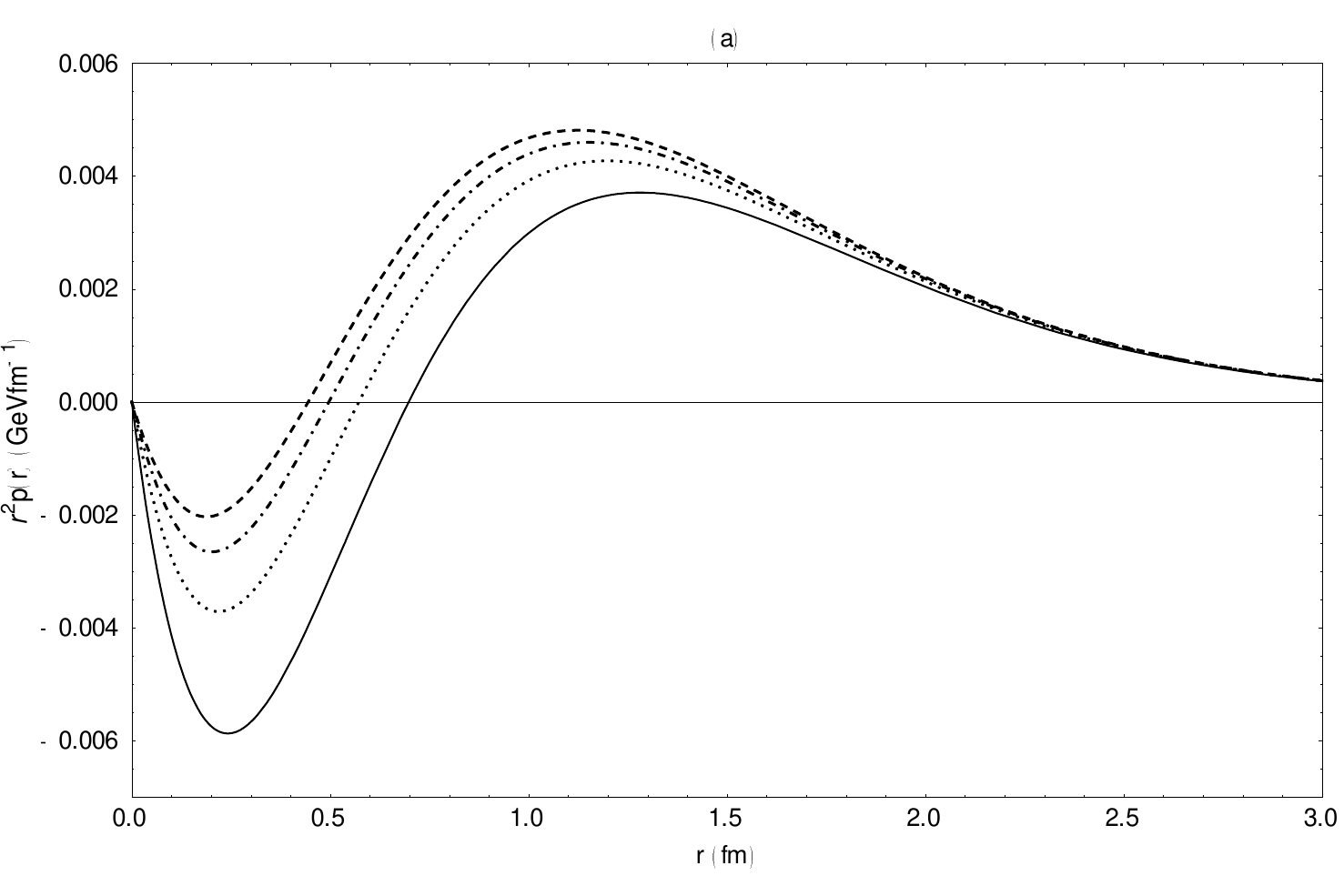}
\includegraphics[scale=0.55]{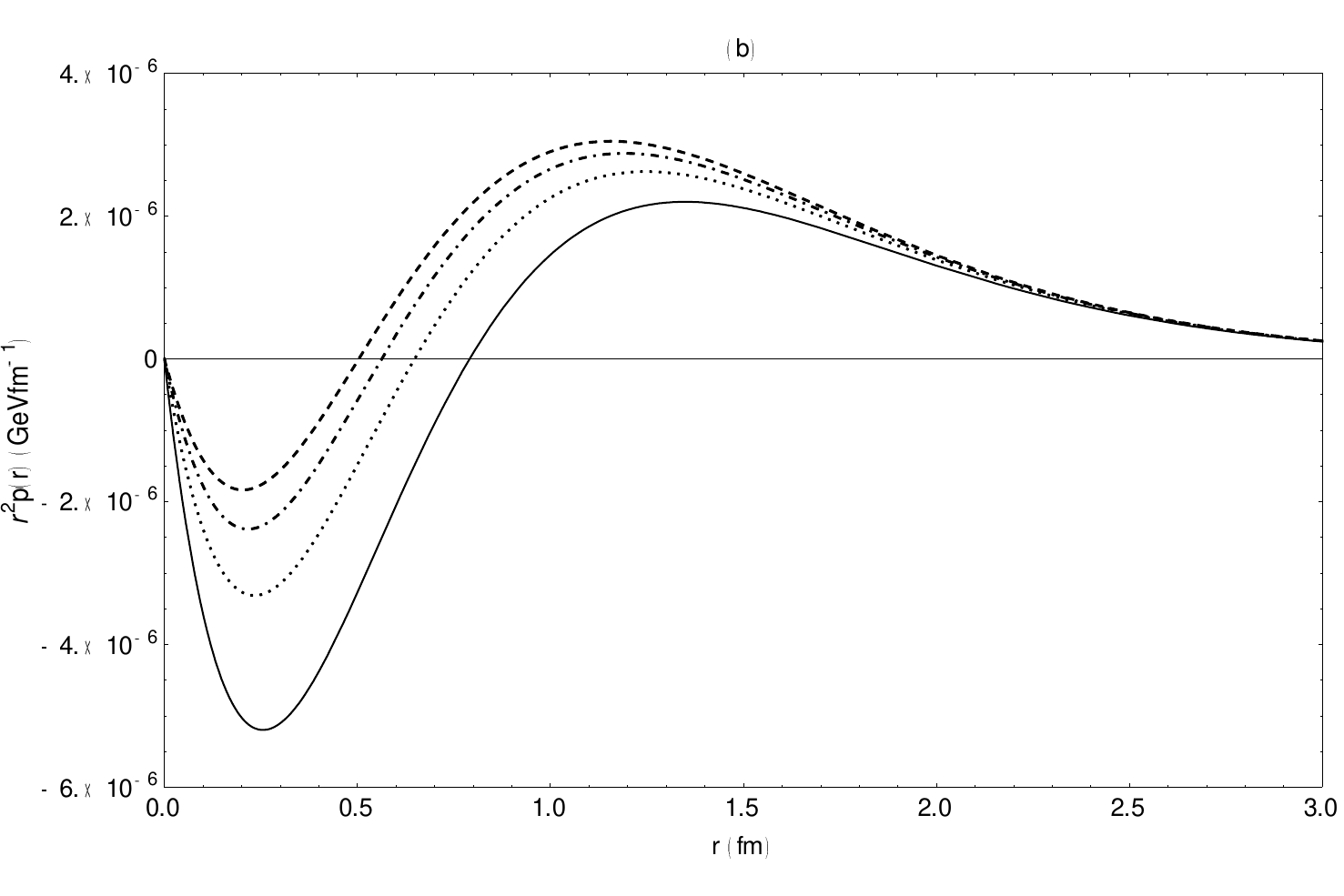}
\includegraphics[scale=0.55]{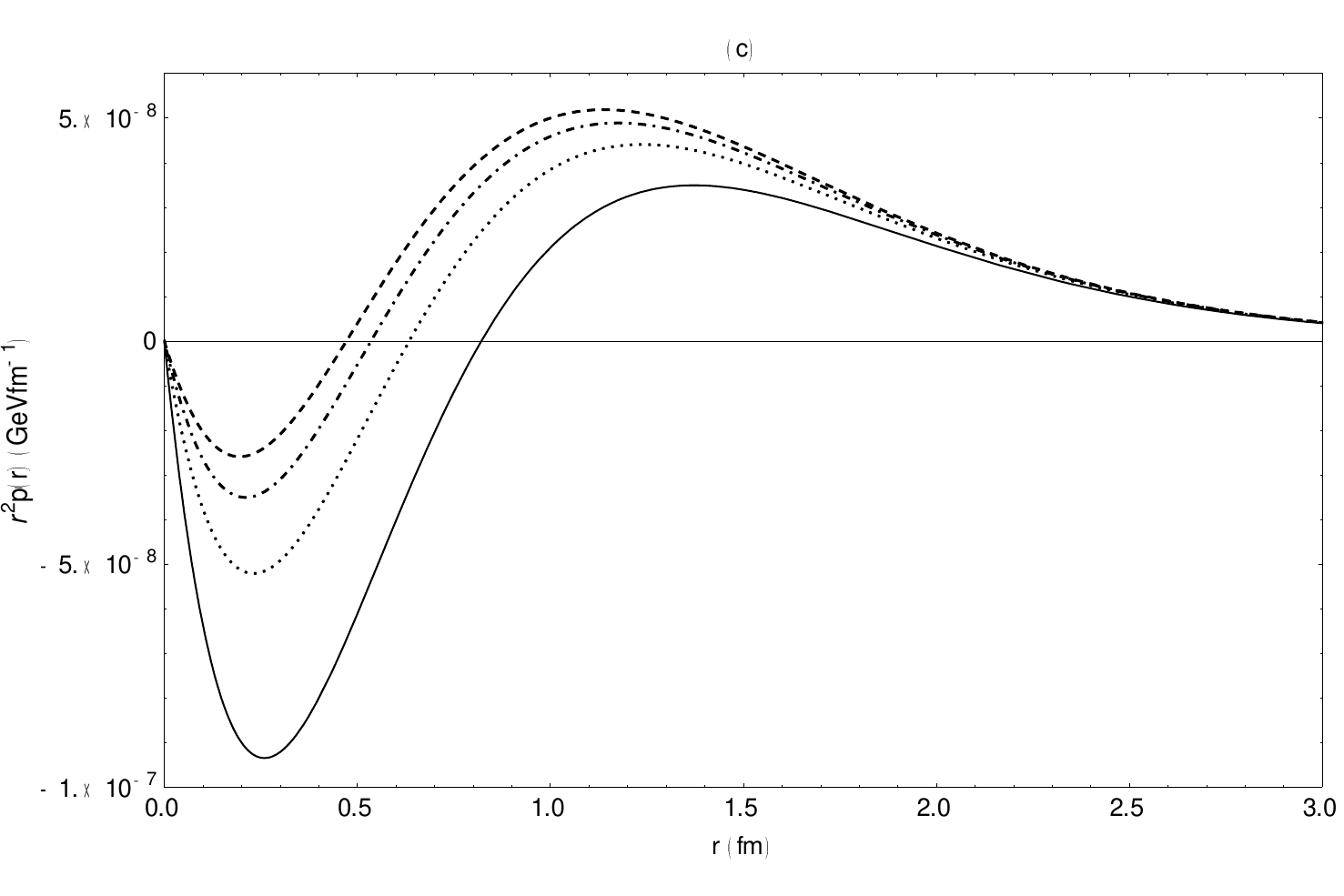}
\includegraphics[scale=0.55]{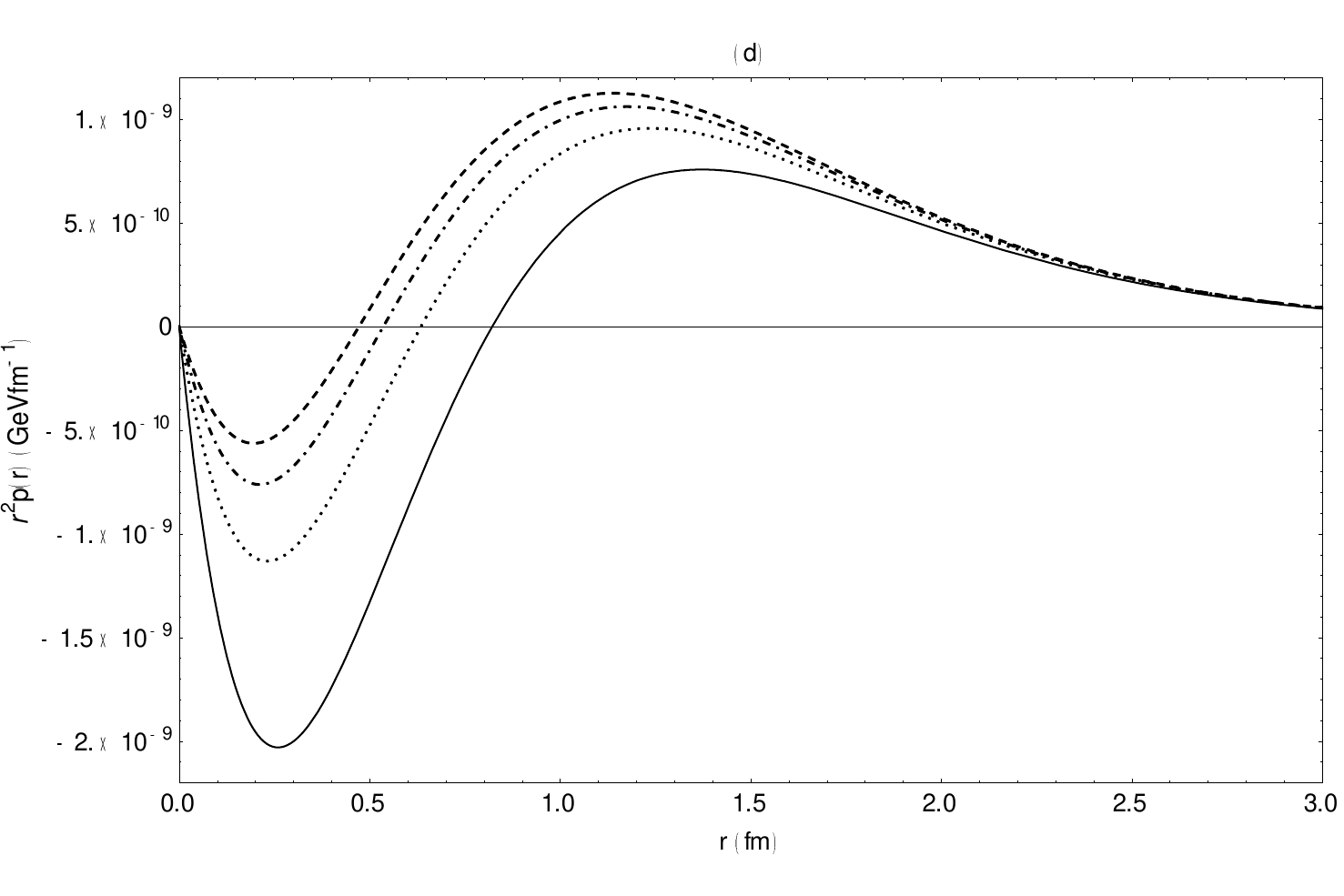}}
\caption{Pressure distribution inside the hadron as a result from the quark interactions versus the radial distance $r$ from the center of the hadron. In all cases, $k=3$ fm$^{-1}$ and $n_f=3$. In (a) $\sqrt{s}=52.8$ GeV, (b) $\sqrt{s}=1800$ GeV, (c) $\sqrt{s}=14000$ GeV, and (d) $\sqrt{s}=95000$ GeV  and for all cases $\sqrt{s_0}=25$ GeV and $Q^2=10^2$ GeV$^2$ (solid), $Q^2=10^4$ GeV$^2$ (dotted), $Q=10^6$ GeV$^2$ (dot-dashed) and $Q=10^8$ GeV$^2$ (dashed). The negative pressure region narrows as the $Q^2$ grows.}
\label{fig:conf_pres_s>s_0}
\end{figure}

Figure \ref{fig:conf_pres_s>s_0} shows the result of Eq. (\ref{eq:model_5}) considering energies above $\sqrt{s_0}=25$ GeV. The negative and positive pressure inversion is due to the phase transition factor. In all cases, the negative pressure region settles down near the center of the hadron (between 0.5 and 1.0 fm), shrinking as $Q^2$ grows. The pairs are now created in the distant periphery ($r>r_0$) and they are not subject to a maximum distance for their creation ($r<r_0$) and, therefore, the confining term of the potential may act to maintain the proton stability.  

The pressure behavior may be a crucial ingredient to understand the total cross section. The interplay between negative and positive pressure distribution inside the hadron allows the following assumption. Suppose that total cross section is the sum of two auxiliary cross sections, one given by the positive pressure ($\sigma_+(s)$) and another one by the negative pressure ($\sigma_-(s)$). Therefore,
\begin{eqnarray}\label{eq:exp1}
\frac{\sigma_-(s)}{\sigma_{tot}(s)}+\frac{\sigma_+(s)}{\sigma_{tot}(s)}=1.
\end{eqnarray}

The relation between these two auxiliary cross sections is responsible for the decreasing and increasing of the total cross section concerning $s_0$. Furthermore, the sum in Eq. (\ref{eq:exp1}) increases or decreases depending on the energy range considered, as can be seen in the total cross section experimental dataset. 

In a geometric point of view, considering $s<s_0$, the dominant auxiliary total cross section is $\sigma_-(s)$ implying the total cross section measures the emptiness of the proton \cite{borcsik_mod_phys_lett_a31_1650066_2016}. The $q\bar{q}$-pairs in this energy regime are created near the center of the proton, being shielded by a negative pressure whose origin is the strong interaction (more precisely, the Coulomb-like term of the potential). On the other hand, when $s>s_0$, then $\sigma_+(s)$ is the dominant part and $\sigma_{tot}(s)$ measures the presence of a well defined structure (or less opaque) \cite{borcsik_mod_phys_lett_a31_1650066_2016}. The $q\bar{q}$-pairs are created now near the hadron periphery, and as the energy grows, the filling up mechanism occurs from the periphery towards the center. The negative pressure region is now at the hadron center.  



\section{Hollowness Effect}\label{sec:hollowness}

The TOTEM \cite{totem_2013} and ATLAS \cite{atlas_2014} experimental results for the differential cross section for $pp$ collisions at $\sqrt{s}=7$ TeV \cite{aaboud_atlas_2016} and $\sqrt{s}=8$ TeV \cite{alkin_phys_rev_d89_091501_2014} seems to indicate a saturation at $b\neq 0$ fm, contrary to the usual theoretical view where the saturation occurs at $b=0$ fm ($b$ is the impact parameter). The inelastic cross section obtained from these dataset, in the impact parameter representation, presents a very smooth hollow at $b=0$ fm. 

There are several papers devoted to explain that unexpected feature \cite{alkin_phys_rev_d_89_091501_2014,dremin_1,dremin_2,anisovich_phys_rev_d_90_074005_2014,troshin_mod_phys_lett_a_31_1650079_2016, troshin_eur_phys_j_a_53_57_2017,broniowski_acta_phys_polon_b10_1203_2017,arriola_phys_rev_d95_074030_2017}.  In particular, one stress here the analysis carried out by Broniowski \textit{et al.} \cite{broniowski_acta_phys_polon_b10_1203_2017}. Based on the Barger-Phillips model \cite{phillips_phys_lett_b_46_412_1973}, the authors of Ref. \cite{broniowski_acta_phys_polon_b10_1203_2017}, assuming a transferred momentum independence on the $\rho$ parameter, introduces a prescription on the scattering amplitude. As a consequence, the hollowness effect emerges being addressed to a quantum origin. In particular, they had pointed out that the real part of the scattering amplitude is responsible for generating the hollowness effect when the real part of the eikonal phase becomes larger than $\pi/2$. 

The theoretical approach presented here is able to furnish the pressure distribution inside the proton, revealing two main results. The first one is the agreement with the experimental result of Ref. \cite{burkert_nature_vol_557_396_2018}. The second one is the prediction of the inversion of negative and positive pressures regions due to the dynamical phase transition occurring in the total cross section at $s_0$. Then, the results possess origin in both the damped confinement potential and in the dynamical phase transition occurring in the total cross section. Therefore, the origin of the pressure distribution is due to the strong interaction acting on the constituents of the proton. 

If the filling up mechanism occurs from the periphery towards the center of the proton for $s>s_0$, then it is plausible to suppose that the hollowness effect may be related with the negative pressure region located near the center of the hadron. As $Q^2$ increases, the negative pressure region shrinks (but seems not to vanish). A non-vanishing real part of the scattering amplitude for $s\rightarrow\infty$ may also be responsible for this result according to the model adopted in Ref. \cite{broniowski_acta_phys_polon_b10_1203_2017}. Of course, the hollowness effect possesses different origins comparing the present-result and those based on the coherence folding of inelasticities of collisions of partonic constituents \cite{broniowski_acta_phys_polon_b10_1203_2017}.

Of course, one cannot expect that pressure inside the proton, specifically the size of the negative pressure region, reflects itself in a hollow region with the same size in the impact parameter space (e.g, in the inelastic overlap function). 

\section{Critical Remarks}\label{sec:critical}

The phase transition occurring inside the hadron, shown by the increasing of the total cross section, $\sigma_{tot}(s)$, as the energy grows, was firstly predicted by Cheng and Wu based on massive electrodynamics \cite{cheng_phys_rev_lett_24_1456_1970,cheng_phys_rev_d1_1064_1970}. However, in that time, unfortunately, it was not claimed as a phase transition. Nonetheless, any treatment of the total cross section should take into account this phase transition \cite{borcsik_mod_phys_lett_a31_1650066_2016,campos_preparation_2018}.

In order to explain a recent experimental result, the model proposed here uses a naive approach: a damped confinement potential to mimic the hadron internal energy. Although naive, the approach used here takes into account a phase transition factor in the pressure calculation and, as the main result, the model is in accordance with the recent measurement of the proton internal pressure \cite{burkert_nature_vol_557_396_2018}. Furthermore, the model predicts the inversion of the pressure regimes for squared energies above $s_0$, where the total cross section achieves its minimum value.  

The results are robust under a wide variety of the parameters values. However, there is a need for the correct experimental determination of some parameters ($k_D$, for instance). In spite of, it is correct to affirm that the negative pressure region is located above the physical root $ r_0 $, for $ s <s_0 $. The constituents near the center of the hadron, $ r <r_0 $, are subject to a Coulomb-like potential. From the experimental point of view, the total cross section decreases in this energy range. Hence, the sum $\sigma_-(s)+\sigma_+(s)$ decreases as $s$ grows, achieving its minimum at $s_0$. As the phase transition takes place, the negative pressure region changes its location towards the hadron center, shrinking as $Q^2$ grows. On the other hand, the constituents placed at some $r>r_0$ feels a positive pressure. In this situation, the sum $\sigma_-(s)+\sigma_+(s)$ increases as $s\rightarrow\infty$. The total cross section is now an increasing function of $s$.

Changes in the hadron pressure regime occur as a result of a phase transition taking place at $s_0$. There is also a phase transition in the confinement potential since it presents a natural decoupling phase in its formulation. However, only the introduction of the phase factor allows the correct pressure regimes inside the hadron.

Considering the hollowness effect, the model proposed suggests that the negative pressure region, emerging for $s>s_0$ near the center of the proton, may be related with the hollow near $b=0$ fm. Of course, the size of the negative pressure region is not the same of the hollow region, as pointed out in the TOTEM results. Possibly, an unknown physical mechanism acts to minimize the negative pressure region effects in this energy regime in the sense that, for small energies, the hollow region is not visible in the impact parameter space. For very high-energies, the hollowness region starts to appear. It is natural to suppose, as pointed out in Ref. \cite{broniowski_acta_phys_polon_b10_1203_2017} (and references therein), that the real part of the scattering amplitude is responsible by this effect.

In conclusion, the model presented here, although naive, is able to furnish an explanation for the recent experimental results. A more sophisticated model taking into account kinematic terms should be considered elsewhere.

\section*{Acknowledgments}

SDC thanks to UFSCar by the financial support.

\end{document}